# Using digital twins for managing change in complex projects

Jennifer Whyte[1,5], Ranjith Soman[2,5], Rafael Sacks[3], Neda Mohammadi[6,1], Nader Naderpajouh[1], Wei Ting Hong[1] and Ghang Lee[4]

[1]University of Sydney, Australia, [2]TU Delft, Netherlands, [3]Technion University, Israel, [4]Yonsei University, Korea, [5]Imperial College London, UK, [6]Georgia Tech, USA

**Abstract**

*Complex systems are not entirely decomposable, hence interdependences arise at the interfaces in complex projects. When changes occur, significant risks arise at these interfaces as it is hard to identify, manage and visualise the systemic consequences of changes. Particularly problematic are the interfaces in which there are multiple interdependencies, which occur where the boundaries between design components, contracts and organisation coincide, such as between design disciplines. In this paper, we propose an approach to digital twin-based interface management, through an underpinning state-of-the-art review of the existing technical literature and a small pilot to identify the characteristics of future data-driven solutions. We set out an approach to digital twin-based interface management and an agenda for research on advanced methodologies for managing change in complex projects. This agenda includes the need to integrate work on identifying systems interfaces, change propagation and visualisation, and the potential to significantly extend the limitations of existing solutions by using developments in the digital twin, such as linked data, semantic enrichment, network analyses, natural language processing (NLP)-enhanced ontology and machine learning.*

## 1. Introduction

Complex projects arise in sectors such as infrastructure, new energy and resources. Systems integration challenges are a significant problem (Whyte & Davies, 2021; Whyte et al., 2022). Yet, there is the potential for advanced digital methods to provide information to engineers and decision-makers that is needed to better manage these changes (Papadonikolaki et al., 2022; Whyte et al., 2016). This is important as complex projects are increasingly delivering cyber-physical systems as interventions into existing natural as well as built environments. As the number of internal and external interfaces grows, new systems integration challenges emerge (Whyte & Davies, 2023). These challenges are particularly salient when late changes are made to designs: even internationally leading engineering firms can find addressing these challenging as they sometimes have unrecognised systemic impacts (Whyte et al., 2016).

Recent work on the digital twin is the enabler of the development of novel methods. While prior research on digital methods for managing change has sought to use both geometric information about adjacencies (Chen & Whyte, 2022; Jacob & Varghese, 2018) and time-series data on design (Gopsill et al., 2016) to identify interfaces and interdependencies, and there is work on systems analyses using a digital twin (Whyte et al., 2019) and change propagation (Giffin et al., 2009), these methods typically have used a limited set of techniques. Building on work on Building Information Modelling (Sacks et al., 2018) and using developments in the digital twin (Mohammadi & Taylor, 2020), there are new opportunities to very significantly enrich and extend such prior analyses including techniques for linked data (Soman et al., 2020), semantic enrichment (Sacks et al., 2020), network analyses (Valentin et al., 2018), and machine learning (Chen & Whyte, 2022; Wang et al., 2022).

This work is timely because projects are becoming larger and more complex moving beyond engineers' ability to understand their complexity using traditional methods. At the same time data provides opportunities, but without the right tools, it can make the problem worse as managers and engineers can become overloaded with too much information. What is lacking are sophisticated data-driven methods to support engineers in rapidly and proactively (ex-ante) identifying relevant interfaces and understanding the systemic impacts of potential changes in design and delivery in large datasets. Given recent advances in work on digital twins, linked data and semantic enrichment there is an opportunity to develop these data-driven methods.

This paper provides an underpinning state of the art of the existing literature, and a small pilot to identify the characteristics of future solutions, and opportunities for research. In section 2 we provide a theoretical background and a review of the existing state-of-the-art in systems approaches to infrastructure, managing change (identification of interfaces and interdependencies, analyses of change propagation), the developments in the digital twin, and new areas in which change analyses become important. In section 3 we then present methods and preliminary work as a basis for new work to provide a step-change new solution to managing change in complex projects by leveraging the digital twin. In section 4 we then present an overview of the research opportunities that we identify. The final section draws some conclusions on the current state-of-the-art and directions for future research.

## 2. Theoretical Background and State-of-the-Art

We build on work to conceptualise and model interdependent large-scale infrastructure systems (Choi et al., 2017), and to understand the propagation of changes within them (Brahma & Wynn, 2023; Giffin et al., 2009). Engineering change has been described as a six-step process, as shown in Figure 1. In practice, we have found that steps 2 and 3 are not well undertaken, and typically consider a very limited range of factors, with potential impacts of the change emergent in stage 6, rather than ahead of the approval of the change.

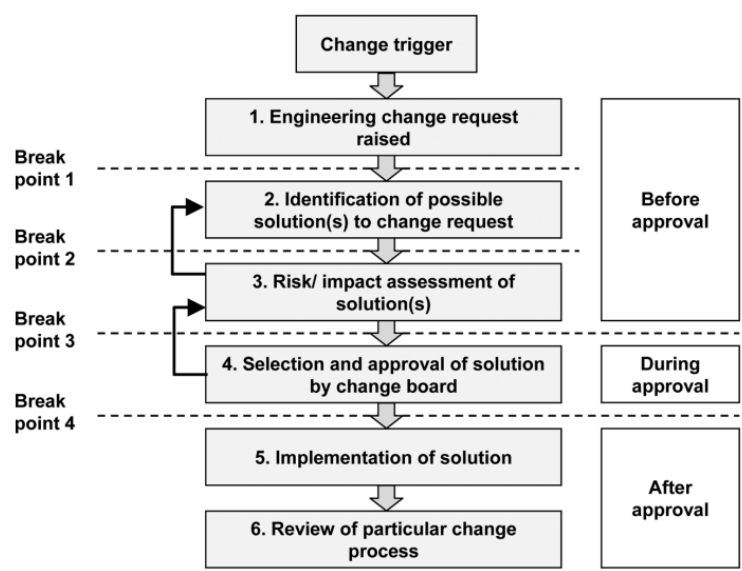

Figure 1: The Engineering Change process, from Hamraz et al. (2013), who adapts from Jarratt et al. (2005)

In complex projects, this process is particularly challenging as major technical interface challenges arise across different engineering disciplines and organisational boundaries, e.g. where accountability and responsibility become unclear, and changes in architecture result from emerging complexities and uncertainties (e.g. between systems with mature and novel technologies projects (Whyte & Davies, 2021)).

## *Systems approaches*

From a systems perspective, managing engineering change is related to the overall architecture as it forms a step in a configuration management process (Ali & Kidd, 2014). It is particularly associated with configuration control, where all changes to configuration items are controlled to ensure the integrity of the overall system. Such techniques, developed in safety-critical sectors such as nuclear and defence, provide a robust framework for managing engineering systems through their design, delivery and operation.

Associated with a systems perspective, there are a variety of existing systems approaches to modelling the structures and dynamics of engineering systems that projects deliver, with a variety of meta-modelling approaches, including model-based systems engineering (MSBE) (Menshenin et al., 2021; Roodt et al., 2020). Accountability and responsibility for systems integrity have been mapped using the systems engineering 'V' diagram, e.g., in relation to requirements (Chen & Jupp, 2023). Process systems engineering has been used to map the engineering interactions through the project delivery process, retrospectively identifying how systems failures arose, e.g., in the case of the Grenfell disaster (Hackitt, 2018).

This systems approach recognises the different degrees of modularity of different architectures, which will create more or less complexity to be managed through a change process, and more or less potential for innovation (Hall et al., 2020). One area of concern should be to reduce complexity at the outset, and another to manage the interfaces where interdependencies persist, so engineering changes can be understood and their impacts controlled.

## *Managing change*

Managing design change, within the wider frameworks, requires the identification of systems interfaces, analyses of change propagation across these and visualisation of outcomes for decision makers. In complex projects, substantial work proceeds through the use of the Design Structure Matrix (DSM) (Browning, 2015) (Also known as an N2 Interface Matrix) as a tool to identify the impacts of change in large engineering designs, using geometric data on connections between components (Chen & Whyte, 2022; Jacob & Varghese, 2018) and process data on co-viewing of different aspects of design (Gopsill et al., 2016) with unrealised opportunities to combine these approaches. This interdependence across interfaces can relate to physical connections, to energy, mass and information flows (de Weck, 2015).

Efforts to analyse and minimise the impact of design changes have been the subject of many studies. One notable theoretical framework is the "patching" proposed by Eastman et al. (1997). Patching is an action that modifies a design as locally as possible to minimise impact while maintaining global integrity with the rest of the design. A range of manual and analytic approaches have been developed, including the application of probabilistic methods to a network of dependencies, as shown in Figure 2.

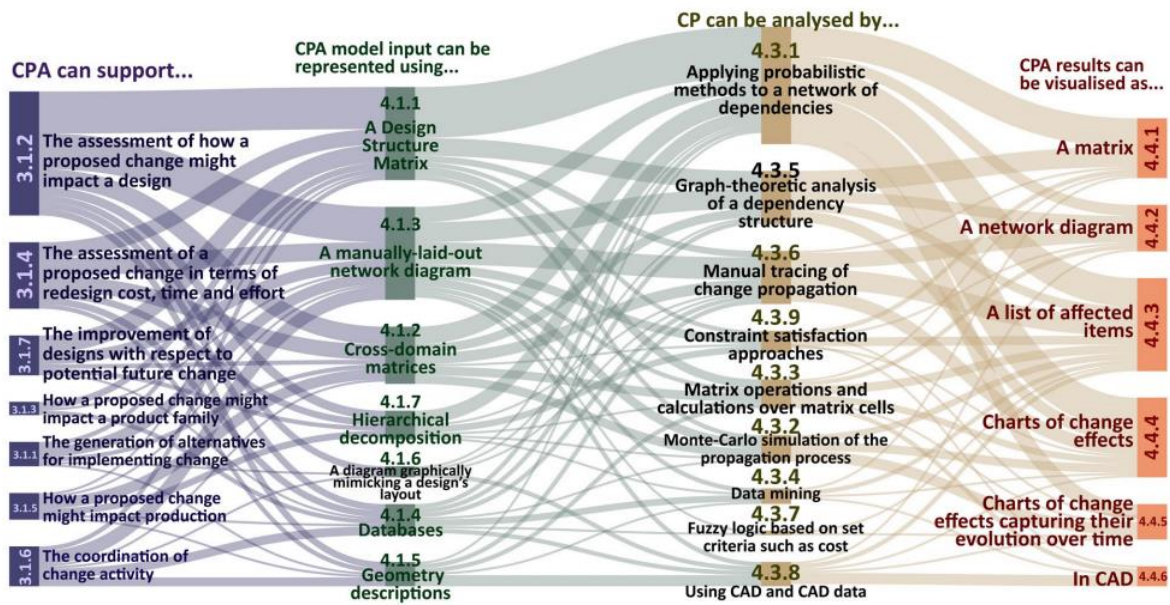

**Figure 2: State of the art on work on change propagation analyses (Brahma & Wynn, 2023).**

## *Developments in the digital twin*

While most work has been on digital twins in operations, recent research advances an emerging area of work on their use in design and construction (Sacks et al., 2020; Tzachor et al., 2022). Research on the digital twin in construction is underpinned by standard ways to describe data (Drogemuller et al., 2021; Farghaly et al., 2024). Within the field of research on AI and design (Allison et al., 2022; Wang et al., 2021), different approaches to identifying these engineering interfaces are emerging.

Link and display heterogeneous data-sets to enable decision-making, with existing work focused on constraints in scheduling data (Soman & Molina-Solana, 2022), and the visualisation of requirements and outcomes in dashboards and indicators, for example, in a construction production control room (Farghaly et al., 2021), and the potential for extension to support an 'interface digital twin'.

## *New areas*

Growing challenges of sustainability and resilience are increasing the extent to which engineering systems, delivered through complex projects, need to be seen as open rather than closed systems. Much of the work on managing change has focused internally within the project boundaries. At the same time, significant advances have been made in recent years with a focus on modelling at the national and regional level, with relatively little connection to the project level. Yet, to tackle systemic issues such as resilience, new approaches are needed to understand external as well as internal interdependencies in ways that inform decisions on projects.

## 3. Methods and Preliminary Work

We conducted preliminary work to scope the need and future technical work to develop digital intelligence to resolve the challenges of systems integration and enable digital and data-driven decision-making using digital twins and analytics. Our modelling and analysis work will underpin the invention of an 'interface digital twin' for systems engineering on major infrastructure projects.

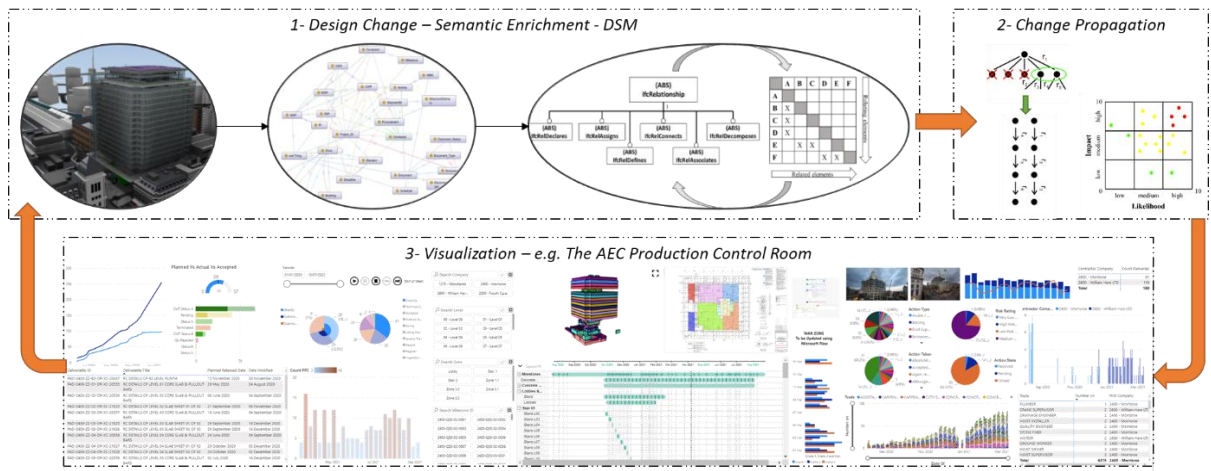

**Figure 3: Pilot to Synthesise existing work on design change, change propagation and the visualization of systemic impacts associated with changes and other factors.**

Figure 3 describes an approach to synthesising existing work on design change, change propagation and the visualization of systemic impacts associated with changes and other factors. Here the ambition is to develop a prospective approach to identifying and representing major interfaces and interdependencies to enable systems engineering and design flexibility on major projects (Farghaly et al., 2024; Farghaly et al., 2021).

Important within this are new approaches that the team has begun to pilot on:

1) *Addressing uncertain and missing data:* The proposed research infers connections in BIM using semantic enrichment through machine learning (Krijnen & Tamke, 2015; Sacks et al., 2017; Wang et al., 2022) and uses emerging methods for generating interface graphs (Ismail et al., 2017). A novel step is the shift from DSM to knowledge graphs, to include semantic elements, beyond the known dependencies, to provide insight on impacts such as change propagation and the organisations that should be involved in resolving interface issues. The idea is to identify relationships between elements from different sub-systems that may be overlooked by designers whose attention is focused on their own domain of expertise, and which thus remain implicit. Learning from graphs derived from previous projects may enable AI routines to enrich this information and make it explicit for use by interface-checking software. DSM is a representative way of dealing with uncertainty, that is missing data before knowledge graphs. To make the step from DSM to knowledge graphs we are taking the opportunity to use *graph data models* to represent systems in graph data models to optimise the configuration of components e.g. building on Khalili and Chua (2013), and to use *multilevel exponential random graph models (MERGM)* to extend the representation of the graph data model of the system into the infrastructure system of systems (Wang et al., 2016; Zappa & Lomi, 2016).

2) *Using the digital twin to inform engineering decisions:* An innovative synthesis of insights from systems engineering, infrastructure design, building information modelling and management and project management to model networks for the identification of major interfaces in the systems architecture. A multi-level analysis approach integrated with natural language processing (NLP)-enhanced ontology and semantic enrichment can efficiently characterise and summarise system-wide design complexities. Such an analysis leverages novel NLP techniques to improve the process of structuring and managing the knowledge presentation involved in changes in complex projects. Innovation is needed to link data-sets, visualising outputs and outcomes to enable proactive decision-making and to take dynamic approaches to address emerging complexity and uncertainty. Developing an interface digital twin is challenging as diverse engineering disciplines use models differently, focusing on time series data (behaviours, performance) or design data (components, interfaces); with project managers using product and work breakdown structures, schedules and process maps to organise delivery.

We have also sought to model the dynamics of systems architecture and analyse these using multi-level networks to enable an understanding of complexities associated with the integration of project components in the wider systems of systems, which provides insights on how to practically include external interdependencies in ways that inform decisions on projects.

## 4. Research agenda for digital twin-based interface management

We anticipate an approach to digital twin-based interface management that extracts design information from information models, evaluates interfaces and interdependencies in complex systems, automates methods and integrates and visualizes the outputs to enable proactive decision-making and validates the approach.

On complex projects the numbers of interfaces and interdependencies can be very large, and there are opportunities to use graphs of these as a scalable non-labular database solution. There is a need for a method for the generation of *interface knowledge graphs* that enable the identification of interfaces between physical, contractual and organisational systems in infrastructure design (Figure 4). Sources of information are BIM models, product and work breakdown structures, drawings, meeting minutes and other available engineering design and construction materials. Missing information on interfaces can be identified through an NLP-enabled ontology of interfaces, BIM log mining and graph inferencing. In doing so, BIM log mining (Jang et al., 2023) and NLP on process knowledge (Hong et al., 2024) are used to generate an enriched interface knowledge graph with knowledge of interface relations from both design and process data. Additionally, we also leverage graph-based inferencing on data from the specific case to infer missing connections in the interface knowledge graph and identify graph vulnerabilities.

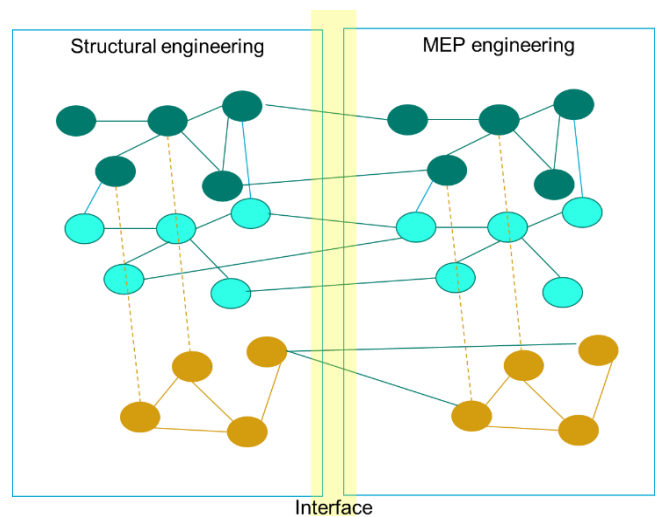

**Figure 4: Interface knowledge graphs with nodes (e.g., design components (physical), design processes (organisational), design requirements (contractual), and links (interdependencies) at interfaces between disciplines).**

To address the potential for change scenarios in the design interdependencies there is a need for generating risk-informed knowledge graphs, namely *risk register knowledge graph*. Risk registers are the core artifacts culminated from tacit and explicit knowledge associated with risk (Khallaf et al., 2018). Connecting the knowledge interface with risk registers provides a feedback mechanism between the risk register and the interface knowledge graph (from) to suggest the potential for reducing complexity and design flexibility based on potential risks (Farghaly et al., 2022). Such risks can be associated with the operation or design of interdependent infrastructure systems (Choi et al., 2017). The feedback mechanism provides a more comprehensive understanding across both knowledge domains by (i) informing risk registers from the interface knowledge graph that connects physical, contractual and organisational interfaces, (ii) informing the interface knowledge graph by the textual and quantitative information associated with risk registers. As a result, there is a chance to prioritise changes and allow design flexibility (as framed by Cardin et al. (2013)) through a risk-informed decision-making process.

Next, the *interface knowledge graphs* and *risk register knowledge graph* are integrated. An illustration of the relationship between the interface knowledge graph and the risk-informed interface knowledge graph is shown in Figure 5. Manual integration will be followed by automated methods such as automated ontology alignment using semantic and structural embeddings of knowledge graphs (Hao et al., 2023). This integration is critical as risks in complex systems are dynamic during the changes in the design interdependencies. Therefore, the layer of information about potential risk scenarios from the risk register knowledge graph combined with the interdependencies from the interface knowledge graph will help identify and manage emerging risks from design changes. Design flexibility is also enabled through the use of causality in selected knowledge graphs to identify change propagation (Germanos et al., 2024).

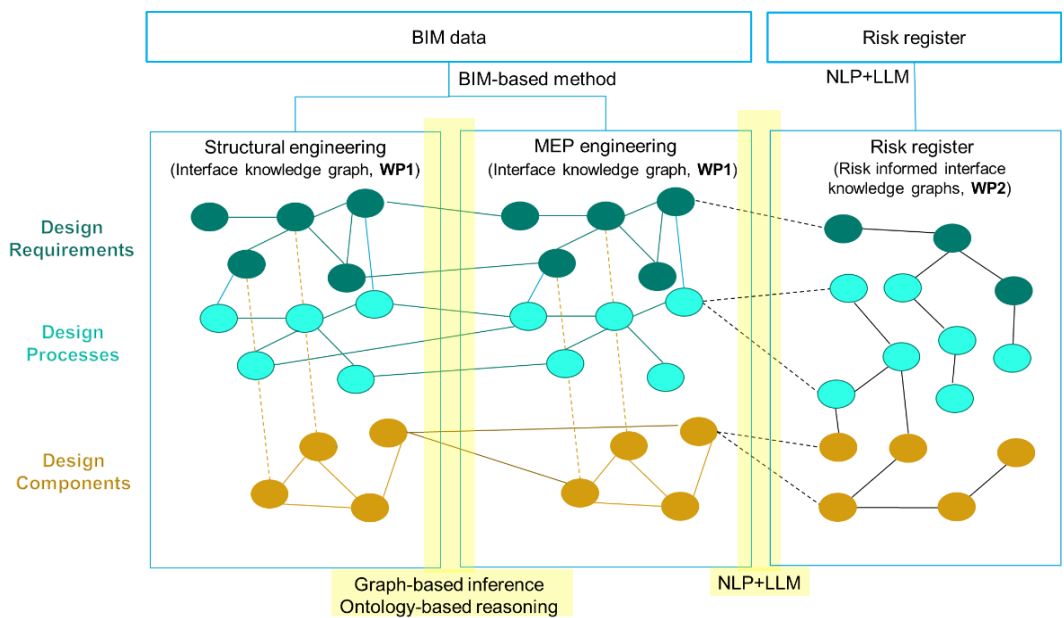

**Figure 5: The relationship between interface knowledge graph and risk-informed interface knowledge graph.**

There is a need for a methodology for automating the extraction and augmentation of BIM data on systems interdependencies. The aim is to enable proactive decision-making by using semantic enrichment and automation in the identification of interfaces, interdependencies and risks in the components and systems architecture of a new project. Potential interface issues and critical changes in projects requiring attention will be highlighted by comparing the graph representations with knowledge graphs, enabling System integration and visualization to further analyse interdependencies of interface issues to inform decision making.

Graphs of interdependencies are not readable to engineers and managers, and hence interpreting the information to inform decision-making is important. The development of advanced visual methods to make information 'human-readable' to decision-makers is crucial to enable the rapid and proactive use of our digital twin-based interface management methods in project decision-making. The dashboards and indicators developed for the interface management methods aforementioned aim to operate like a search tool for engineering design data, enabling an 'interface digital twin', with information in a digital twin representation that attracts engineers' and designers' attention to the areas of most likely concern in a system configuration.

These methods developed in the lab can be validated through industry trials and benchmarking, allowing engineers and designers to engage directly with the tool, and also get their feedback on pathways to impact and the potential for collaborating to extend this basic research to develop practical use cases in areas of current need (e.g. water, housing and energy).

## 5. Conclusions

We set out an approach to digital twin-based interface management and an agenda for research on advanced methodologies for managing change in complex projects. Recent developments in the digital twin make possible a new generation of techniques for managing change on complex projects. This is important because the scale and complexity of construction and infrastructure projects is growing, and we are becoming more aware of their interfaces and interdependencies, within the project and across project boundaries. The contribution of this paper is to identify opportunities to combine digital-twin-based methods with advances, e.g. in linked-data, ML and semantic enrichment, to significantly advance the ability to address interdependencies at interfaces in complex projects.

First, a step-change is needed to identify and visualise uncertain and missing information and give insight into the nature of interdependencies. This addresses the significant limitation of existing approaches is the missing connection data within engineering design models and Building Information Modelling (BIM).

Second, a step-change is needed to inform decisions before they are made, to enable proactive decision-making and to take dynamic, rather than static, approaches to addressing emerging complexity and uncertainty. This addresses a significant limitation in existing work, in which the comprehensive analyses of systems and their delivery in existing approaches take too long to inform the time-critical decisions that engineers and managers make on projects. Here,

Our preliminary work suggests that the needed step-changes are ambitious, not yet achieved, but now possible. They need new digital methods and algorithms using emerging approaches to achieve semantic enrichment via NLP technology and design flexibility in the modelling and analyses of systems in major infrastructure projects to manage design change.

The agenda we set out includes the need to integrate work on identifying systems interfaces, change propagation and visualisation, and the potential to significantly extend the limitations of existing solutions by using developments in the digital twin, such as linked data, semantic enrichment, network analyses, natural language processing (NLP)-enhanced ontology and machine learning.

We suggest some directions for future research, both to achieve the step-change and beyond. For example, Scholars of BIM and digital twins can build on this work to develop methods to integrate the learnings on interface as suggestions into design phase models helping the creation of interface knowledge graphs for future projects. This would be an essential step for distributed models in the infrastructure and modularise design processes in complex projects. Scholars of change management can build on this work (Ahn et al., 2017), both to develop new network-based approaches that extend existing matrix methods (such as the DSM) and also to address the significant opportunities to develop a better ex-ante understanding of interfaces and interdependencies within projects (internal interfaces between systems and components) and across their boundaries (addressing larger-scale issues of resilience). Scholars of project management and design can use such methods to improve the flexibility of design in complex projects, where there are emergent complexities as well as those identified at the outset where technologies are developing at different rates, for example in airport projects, where baggage handling systems are updated on relatively short timescales, such that it is important to understand interdependencies in order to leave flexibility in the design to accommodate new systems.

The new approaches we outline provide methods for engaging with large scale, heterogeneous data, to understand systemic consequences of changes and enable better real-time decisions on projects. New research can develop approaches to use emerging techniques to better manage risks across portfolios and programs of projects.